\documentclass[showpacs,preprintnumbers,amsmath,amssymb,superscriptaddress,pra,twocolumn]{revtex4}

\include{amsmath}
\usepackage{graphicx}
\usepackage{subfigure}
\usepackage{amsmath}
\usepackage{times}

\newcommand{\M}{\mathrm{M}}
\newcommand{\A}{\mathrm{A}}

\newcommand{\C}{\mathrm{C}}
\newcommand{\D}{\mathrm{D}}
\newcommand{\B}{\mathrm{B}}
\newcommand{\LL}{\mathrm{L}}
\newcommand{\RR}{\mathrm{R}}

\newcommand{\ccr}{\hat{c}^\dagger}
\newcommand{\cde}{\hat{c}}
\newcommand{\acr}{\hat{a}^\dagger}
\newcommand{\ade}{\hat{a}}

\newcommand{\xhat}{\hat{x}}

\newcommand{\xihat}{\hat{\xi}}
\newcommand{\dhat}{\hat{d}}
\newcommand{\bhat}{\hat{b}}

\newcommand{\du}{d}
\newcommand{\im}{i}
\newcommand{\ex}{e}

\newcommand{\therm}{\mathrm{th}}
\newcommand{\Ar}{\mathrm{A}_\mathrm{r}}
\newcommand{\Ab}{\mathrm{A}_\mathrm{b}}
\newcommand{\Br}{\mathrm{B}_\mathrm{r}}
\newcommand{\Bb}{\mathrm{B}_\mathrm{b}}
\newcommand{\Cb}{\mathrm{C}_\mathrm{b}}
\newcommand{\Db}{\mathrm{D}_\mathrm{b}}
\newcommand{\bb}{\mathrm{b}}
\newcommand{\rr}{\mathrm{r}}

\begin{document}

\title[Proposal for entangling remote micromechanical oscillators via optical measurements]
{Proposal for entangling remote micromechanical oscillators via optical measurements}
\author{K. B{\o}rkje}
\author{A. Nunnenkamp}
\author{S. M. Girvin}
\affiliation{Departments of Physics and Applied Physics, Yale University, New Haven, Connecticut 06520, USA}

\date{Received \today}
\begin{abstract}
We propose an experiment to create and verify entanglement between remote mechanical objects by use of an optomechanical interferometer. Two optical cavities, each coupled to a separate mechanical oscillator, are coherently driven such that the oscillators are laser cooled to the quantum regime. The entanglement is induced by optical measurement and comes about by combining the output from the two cavities to erase which-path information. It can be verified through measurements of degrees of second-order coherence of the optical output field. The experiment is feasible in the regime of weak optomechanical coupling. Realistic parameters for the membrane-in-the-middle geometry suggest entangled state lifetimes on the order of milliseconds.
\end{abstract}
\maketitle

Despite the tremendeous success of quantum mechanics at explaining the behaviour of the microscopic world, many people have found it uncomfortable that macroscopic objects should also obey the laws of quantum mechanics. This goes back to the founders of quantum mechanics, and is the origin of the famous Schr{\"o}dinger cat thought experiment \cite{Leggett2005Science}. However, if all time evolution is unitary according to the Schr{\"o}dinger equation, as in Everett's relative-state interpretation \cite{Everett1957RMP}, there is nothing that forbids counterintuitive phenomena such as superpositions of macroscopically distinct states \cite{Leggett1980ProgTheorPhys}. It should therefore be possible to observe quantum effects on arbitrarily large scales if the system is adequately shielded from environmentally induced decoherence. Conversely, experiments might reveal the existence of unknown sources of decoherence \cite{Ghirardi1986PRD,Penrose1996GenRelGrav}, which limit quantum mechanics to small scales by causing an objective wavefunction collapse and creating a quantum-classical boundary. 

Cavity optomechanics, where mechanical oscillators are coupled to light or microwaves \cite{Kippenberg2008Science,Marquardt2009Physics,Favero2009NatPhot}, is a promising field for experimental tests of quantum mechanics at large scales. The motion of a micromechanical object can be cooled via radiation pressure forces \cite{Cohadon1999PRL,Gigan2006Nature,Kleckner2006Nature,Arcizet2006Nature,Schliesser2006PRL,Teufel2008PRL,Rocheleau2010Nature,Thompson2008Nature}. One group has already reported to have reached the quantum regime by resolved sideband cooling \cite{Teufel2011}, and others are expected to soon follow. An interesting future direction for experiments would be to detect quantum entanglement \cite{Horodecki2009RMP} between a micromechanical oscillator and another system, such as an optical cavity field mode \cite{Vitali2007PRL} or a second mechanical oscillator \cite{Mancini2002PRL,Hartmann2008PRL,Ludwig2010PRA,Pirandola2006PRL}. 

In this Letter, we propose a new and promising route to create and verify entanglement between two remote mechanical oscillators. Our idea is based on an optomechanical interferometer where two optical cavities, each coupled to a separate mechanical oscillator, are coherently driven in parallel. The drive frequency is chosen such that the oscillators are cooled close to the motional ground state. In short, entanglement is achieved by combining the optical output from the two cavities to erase which-path information. The mechanical objects in our setup have no direct interaction, but are projected onto an entangled state through optical measurements, in contrast to various earlier proposals \cite{Mancini2002PRL,Hartmann2008PRL,Ludwig2010PRA}. A scheme related to ours involving laser pulses was discussed in Ref.~\cite{Pirandola2006PRL}. Similar approaches have been successfully applied to entangle remote atomic ensembles \cite{Chou2005Nature,Julsgaard2001Nature} as well as individual trapped ions \cite{Moehring2007Nature}.  We also present a new and feasible way of detecting the entanglement. The experiment we propose should be realizable with present day technology in the regime of weak optomechanical coupling. For the membrane-in-the-middle geometry \cite{Thompson2008Nature}, we estimate that the entangled states can have decoherence times on the order of milliseconds. This could have relevance to quantum information and communication technologies \cite{Duan2001Nature,Kimble2008Nature}.

Consider the standard optomechanical system where the position of a mechanical oscillator modulates the frequency of an optical cavity mode. We discuss a single cavity first and move on to a system with two cavities later. We will have the membrane-in-the-middle setup \cite{Thompson2008Nature} in mind, but our discussion can apply to many other realizations. The photon and phonon annihilation operators will be denoted by $\ade$ and $\cde$, respectively. The interaction Hamiltonian is $\hat{H}_\mathrm{int} = \hbar g \xhat \acr \ade$ where $g$ is a  coupling constant. The mechanical position operator is $\xhat = x_\mathrm{zpf} (\cde + \ccr)$, with $x_\mathrm{zpf}$ being the size of the zero point fluctuations. Input-output theory \cite{Collett1984PRA,Clerk2010RMP} leads to the quantum Langevin equations
\begin{eqnarray}
  \label{eq:quantLangevin}
\dot{\ade} & = & - \left(\frac{\kappa}{2} + \im \omega_\C \right) \ade - \im g \xhat \ade + \sum_i \sqrt{\kappa_i} \, \ade_{\mathrm{in},i}  \ , \\
  \dot{\cde} & = & - \left(\frac{\gamma}{2} + \im \omega_\M \right) \cde - \im g x_\mathrm{zpf} \acr \ade + \sqrt{\gamma} \, \cde_{\mathrm{in}} \ . \nonumber
\end{eqnarray}
The bare mechanical and optical frequencies are $\omega_\M$ and $\omega_\C$. The mechanical oscillator is coupled to a thermal bath, resulting in a nonzero linewidth $\gamma$ and a fluctuating force determined by the operator $\cde_{\mathrm{in}}$. For systems where the mechanical quality factor $\omega_\M/\gamma$ is high, the Markov approximation gives $[\cde_{\mathrm{in}}(t) ,\ccr_{\mathrm{in}}(t') ] = \delta(t-t')$ and $\langle \ccr_{\mathrm{in}}(t) \cde_{\mathrm{in}}(t') \rangle = n_\therm \delta(t-t')$. Here, $n_\therm \approx k_\mathrm{B} T/\hbar \omega_\M$, where $T$ is the bath temperature. The optical linewidth is $\kappa = \sum_i \kappa_i$, where $\kappa_i$ is the decay rate through decay channel $i$. We imagine a two-sided cavity with a left ($\LL$) and a right ($\RR$) input port, and assume that the cavity is driven from the left. The optical input modes then take the form $\ade_{\mathrm{in},\LL} = \ex^{-\im \omega_\D t} \bar{a}_\mathrm{in} + \xihat_\LL(t)$ and $\ade_{\mathrm{in},\RR} = \xihat_\RR(t)$, where $\bar{a}_\mathrm{in}$ is a constant, $\omega_\D$ is the drive frequency, and the vacuum noise operators $\xihat_i$ obey $[\xihat_i(t),\xihat^\dagger_j(t')] = \langle \xihat_i(t) \xihat_j^\dagger(t') \rangle = \delta_{ij} \delta(t-t')$. We define the detuning between the drive and the mean cavity frequency as $\Delta = \omega_\D - \omega_\C - g\langle \hat{x} \rangle$.

The cavity mode operator can be written as a sum of a mean and a fluctuating part, $\ade(t) = \ex^{-\im \omega_\D t}(\bar{a} + \dhat(t))$, where $|\bar{a}|^2$ is the mean number of photons in the cavity. Here, we focus on the situation $\langle \dhat^\dagger \dhat \rangle \ll |\bar{a}|^2$. In that case, Eqs.~(\ref{eq:quantLangevin}) can be linearized and solved analytically. The effective coupling between the optical and mechanical fluctuations is given by $\alpha = g x_\mathrm{zpf} \bar{a}$. In the regime of weak coupling $|\alpha| \ll \kappa$ and for negative detuning $\Delta$, the mechanical oscillator is approximately in a thermal state, but with renormalized parameters compared to the case of $g=0$. The frequency $\tilde{\omega}_\M$ is shifted from its bare value due to the interaction with the optical field, often referred to as the optical spring effect \cite{Braginsky1999PhysLettA}. The effective linewidth $\tilde{\gamma} = \gamma + \gamma_\mathrm{opt}$ is now a sum of the bare value and a contribution $\gamma_\mathrm{opt}$ due to the optomechanical coupling. The latter is positive when the detuning $\Delta$ is negative, leading to line broadening. The effective mean phonon number is the weighted sum $n_\M = (\gamma n_\therm + \gamma_\mathrm{opt} n_\mathrm{opt})/\tilde{\gamma}$, where $n_\mathrm{opt}$ is a measure of the effective temperature of the radiation pressure shot noise \cite{Clerk2010RMP,Marquardt2007PRL,Wilson-Rae2007PRL}. Below, we will set the detuning to $\Delta = -\omega_\M$, which is optimal for cooling, and assume that $\gamma_\mathrm{opt} \gg \gamma$ and $n_\mathrm{opt} = \kappa^2 /(4\omega_\M)^{2} < 1$. Note however that our results are also of interest when $n_\mathrm{opt} > 1$ (see Ref.~\cite{supplementary}).

The optical output mode on the right-hand side of the cavity is $\bhat(t) = \sqrt{\kappa_\RR} \ade(t) - \xihat_\RR(t)$. The optomechanical interaction leads to sidebands in the output which are displaced from the drive frequency $\omega_\D$ by the mechanical frequency $\omega_\M$. The sidebands originate from Raman scattering off the mechanical oscillator, where a photon gains (loses) energy by destroying (creating) one phonon. We will imagine that photons at the drive frequency can be completely filtered away and that the two sidebands can be measured independently. We will refer to the output at frequency $\omega_\D \pm \omega_\M$ as the blue (red) sideband, and denote the output mode filtered around this frequency by $\bhat_\mathrm{b(r)}(t)$. See the Supplementary Material \cite{supplementary} for details. The width of the sidebands are given by the mechanical linewidth $\tilde{\gamma}$, so we assume the filter bandwidth $\lambda$ to obey $\tilde{\gamma} \ll \lambda \ll \omega_\M$. The ratio between the output fluxes of red and blue photons is $n_\mathrm{opt}(n_\M + 1)/(n_\mathrm{opt} + 1)n_\M$. 

A filter that removes the carrier photons at $\omega_\D$ and splits the red and blue photons into different spatial modes might pose a technical challenge. One reason is that the mechanical frequency is typically in the kHz-MHz range. However, there are also experimental setups with $\omega_\M$ in the GHz range \cite{Eichenfield2009Nature_2,Carmon2007PRL,Ding2011ApplPhysLett}. Another reason is that the ratio between the fluxes of blue and carrier photons, given by $4 (g x_\mathrm{zpf}/\kappa)^2 n_\M$, is very small in the weak coupling limit. An alternative and more feasible way to achieve the filtering is through heterodyne photodetection, where the blue and red sidebands can easily be distinguished in the Fourier domain. This is discussed in detail in the Supplementary Material \cite{supplementary}.

We define the degrees of second-order coherence \cite{Loudon2003Book}
\begin{equation}
  \label{eq:g2def}
  g^{(2)}_{j|i}(\tau) = \frac{\langle \bhat^\dagger_i(t) \bhat^\dagger_j(t+\tau) \bhat_j(t+\tau) \bhat_i(t) \rangle}{\langle \bhat^\dagger_i(t) \bhat_i(t) \rangle \langle \bhat^\dagger_j(t) \bhat_j(t) \rangle} \ , 
\end{equation}
where  steady state has been assumed, $\tau > 0$, and the indices $i$ and $j$ denote either red (r) or blue (b). We find that $g^{(2)}_\mathrm{r|r}(\tau) = g^{(2)}_\mathrm{b|b}(\tau) = 1 + \ex^{-\tilde{\gamma} \tau}$. This is what one would expect for thermal radiation seen through a Lorentzian filter of width $\tilde{\gamma}$ \cite{Loudon2003Book}. The photon statistics of the red and blue sidebands is that of thermal radiation simply because the mechanical oscillator is approximately in a thermal state. More interestingly, the cross-correlators become 
\begin{eqnarray}
  \label{eq:g2SingleCavity}
  g^{(2)}_\mathrm{b|r}(\tau)  & = & 1 + \frac{n_\M + 1}{n_\M} \ex^{-\tilde{\gamma} \tau} \ , \\
  g^{(2)}_\mathrm{r|b}(\tau)  & = & 1 + \frac{n_\M}{n_\M + 1} \ex^{-\tilde{\gamma} \tau} \ , \notag
\end{eqnarray}
when restricting the delay time such that $\tau \gg \kappa^{-1} , \ \lambda^{-1}$ \cite{supplementary}. The result (\ref{eq:g2SingleCavity}) has a simple physical explanation. If the effective phonon number $n_\M$ is less than 1, the mechanical oscillator spends most of the time in the ground state $|0\rangle$. It can gain energy and reach the excited state $|1\rangle$ by the creation of a red photon. However, it is bound to return to the ground state quickly, through the creation of a blue photon. This means that conditioned on the detection of a red photon, the probability of detecting a blue photon is high, such that $g^{(2)}_\mathrm{b|r}(\tau)$ becomes large. The opposite is not the case. For $n_\M < 1$, once a blue photon is detected, it probably means that the oscillator is now in the ground state $|0\rangle$ and detection of a red photon is not particularly likely. In fact, in the limit $n_\M \rightarrow 0$, having detected a blue photon does not change the probability of detecting a red one, such that $g^{(2)}_\mathrm{r|b}(\tau) \rightarrow 1$. Note that for sufficiently small $n_\M$, the classical inequality $[g^{(2)}_\mathrm{b|r}(\tau)]^2 \leq g^{(2)}_\mathrm{r|r}(0) g^{(2)}_\mathrm{b|b}(0) $ \cite{Loudon2003Book} can be violated. This happens when $|\langle \bhat_\mathrm{r}^\dagger(t) \bhat_\mathrm{b}^\dagger(t+\tau)\rangle|^2 > \langle \bhat_\mathrm{r}^\dagger(t) \bhat_\mathrm{r}(t) \rangle \langle \bhat_\mathrm{b}^\dagger(t) \bhat_\mathrm{b}(t) \rangle $ and means that the fields $\bhat_\mathrm{b}$ and $\bhat_\mathrm{r}$ are entangled. This is a necessary requirement for detecting entanglement between mechanical oscillators in the setup that we discuss next.
\begin{figure}[htbp]
\begin{center} 
\includegraphics[width=.47\textwidth]{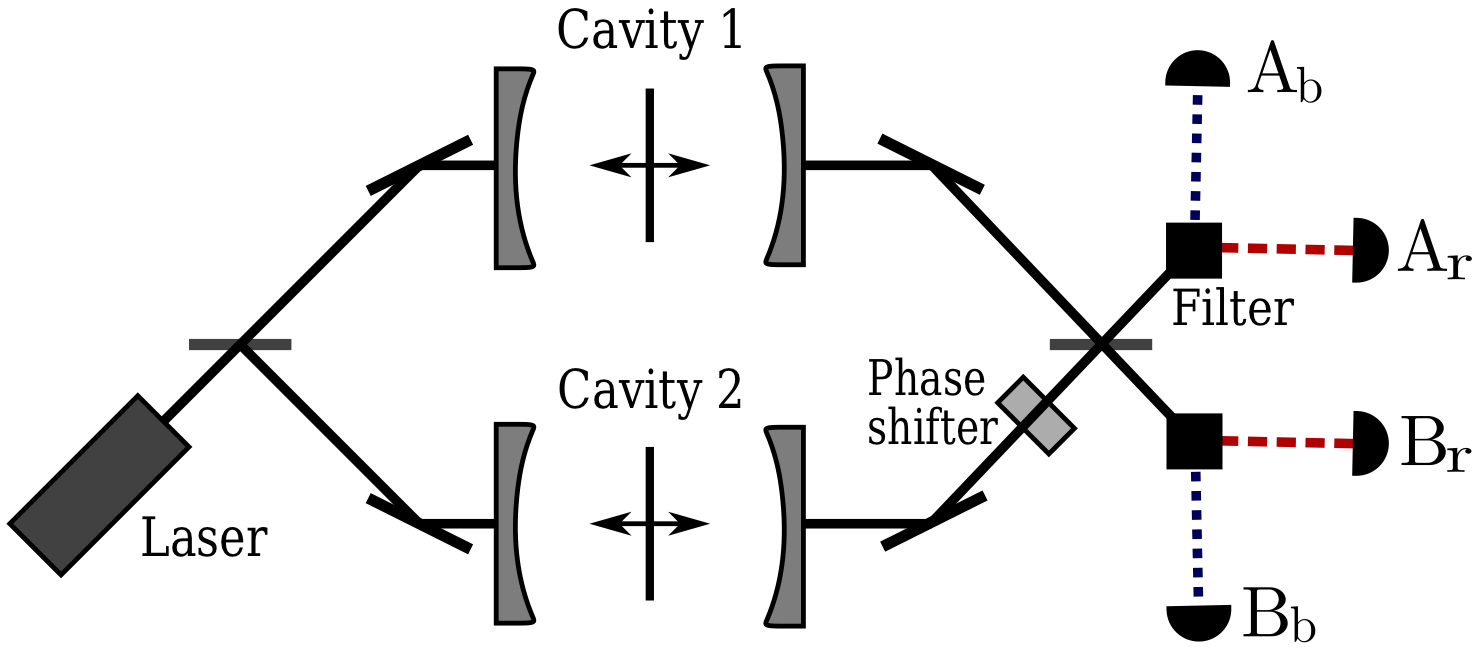}
\caption{(Color online) Schematic view of our proposed experimental setup. Combining the output from the two cavities on a beam splitter can create entanglement between the mechanical oscillators. This can be verified by measuring the photon statistics of the red (dashed) and the blue (dotted) sidebands. The sideband filtering can also be achieved through heterodyne photodetection.}
\label{fig:setup}
\end{center}
\end{figure}

We now move on to the main part of this Letter and study the setup presented in Fig.~\ref{fig:setup}. We consider two optical cavities, 1 and 2, each coupled to a mechanical oscillator. For simplicity, the cavities are assumed to be identical, such that $\omega_\mathrm{C,1}  = \omega_\mathrm{C,2} \equiv \omega_\C$, $\kappa_1 = \kappa_2 \equiv \kappa$, etc. We imagine that $\kappa_\RR \gg \kappa_\LL$, such that most photons leave the cavity through the mirror on the right. Furthermore, we assume that the optomechanical coupling constants are equal. We do however allow for the two mechanical oscillator frequencies to differ, defining $\omega_\mathrm{M,1} = \omega_\M$ and $\omega_\mathrm{M,2} = \omega_\M + \delta$, but we restrict the difference to $|\delta| \ll \kappa, \omega_\mathrm{M}$. The cavities are driven in parallel by a laser optimally detuned to $\Delta = -\omega_\mathrm{M}$, such that cooling to the quantum limit is achieved for both oscillators. The right-hand side output from the two cavities are combined on a 50:50 beam splitter and then filtered into red and blue sidebands, as discussed above. The red photons are detected at either the photomultiplier $\mathrm{A}_\mathrm{r}$ or $\mathrm{B}_\mathrm{r}$, whereas the blue photons are detected at photomultiplier $\mathrm{A}_\mathrm{b}$ or $\mathrm{B}_\mathrm{b}$. 

Let us first discuss how this setup can lead to entanglement between the two mechanical oscillators. Assume that the two oscillators are identical and in the state $|0,0\rangle$, i.e., zero phonons in both. The detection of a red sideband photon means that one mechanical excitation, a phonon, has been created. However, the information on which cavity the photon came from has been erased. This means that, conditioned on detecting a red photon at either $\mathrm{A}_\mathrm{r}$ or $\mathrm{B}_\mathrm{r}$, the wavefunction of the oscillators has now collapsed onto the superposition $(|1,0\rangle + \ex^{\im \theta} |0,1\rangle)/\sqrt{2}$. This is an entangled state, even though it contains only one phonon \cite{vanEnk2005PRA,Cunha2007ProcRoySocA}. The phase $\theta$ depends on whether the photon was detected at $\Ar$ or $\Br$, and other factors such as optical path length differences. While this simplified discussion provides insight on how entanglement is created, the thermal baths must of course be taken into account.  

It is interesting to examine the degrees of second-order coherence defined in Eq.~(\ref{eq:g2def}) for this setup. Now we need to specify not only photon ``color'', but also which detectors we refer to. Taking $g^{(2)}_{\Ab|\Ar}(\tau)$ as an example, it is instructive to express this as
\begin{equation}
  \label{eq:condonAr}
g^{(2)}_{\Ab|\Ar} (\tau) = \frac{\langle \bhat_{\Ab}^\dagger(t') \bhat_{\Ab}(t') \rangle_{\Ar}}{\langle \bhat_{\Ab}^\dagger(t') \bhat_{\Ab}(t') \rangle} \ ,
\end{equation}
defining $t' = t + \tau$. Here, the expectation value in the denominator is the photon flux at detector $\mathrm{A}_\mathrm{b}$ with respect to the state $\rho_\mathrm{ss}$, which is the steady state density matrix in the absence of measurements. In the state $\rho_\mathrm{ss}$ there is obviously no entanglement between the mechanical oscillators. On the other hand, the expectation value in the numerator is defined by $\langle \hat{O}(t') \rangle_{\Ar} = \mathrm{Tr}(\hat{O} \tilde{\rho}(t'))$, where, formally, $\tilde{\rho}(t') = \ex^{{\cal L}(t'-t)} \bhat_{\Ar} \rho_\mathrm{ss} \bhat^\dagger_{\Ar}/\mathrm{Tr}(\bhat_{\Ar} \rho_\mathrm{ss} \bhat^\dagger_{\Ar} )$. This is the time-dependent density matrix conditioned on the detection of a red photon at $\Ar$ at time $t$. The Liouvillian ${\cal L}$ is that of the free evolution of the total system. In the state $\tilde{\rho}(t') $, entanglement between the remote oscillators can occur.

With our assumptions about the cavity and oscillator parameters, the effective phonon number $n_\M$ and linewidth $\tilde{\gamma}$ of the two oscillators will be approximately equal, such that we can drop indices on these quantities. For $\tau \gg \kappa^{-1},\lambda^{-1}$, we find that \cite{supplementary}
\begin{eqnarray}
  \label{eq:g2Interferometer}
  g^{(2)}_{\Ab|\Ar}(\tau) & = & 1 + \frac{n_\M + 1}{n_\M} \ex^{-\tilde{\gamma} \tau} \cos^2(\delta \tau/2 + \phi) \ , \\
  g^{(2)}_{\Bb|\Ar}(\tau) & = & 1 + \frac{n_\M + 1}{n_\M} \ex^{-\tilde{\gamma} \tau} \sin^2(\delta \tau/2 + \phi) \ , \notag
\end{eqnarray}
where the phase $\phi$ depends on path length differences. One could make this phase adjustable by introducing a phase shift in one of the arms of the interferometer, as suggested in Fig.~\ref{fig:setup}. The interference effect in Eq.~(\ref{eq:g2Interferometer}) can be understood classically for large $n_\M$. It simply means that the red light phase difference between the two cavity outputs at time $t$ is related to the blue light phase difference at $t+\tau$ because the mechanical oscillators have a well defined phase difference for times $\tau \lesssim \tilde{\gamma}^{-1}$. This phase difference becomes time dependent if the mechanical frequency difference $\delta$ is nonzero. From a quantum mechanical point of view, the detection of a red photon creates a mechanical superposition with a definite phase $\theta$. The subsequent blue photon will be superposed between the upper and lower arms of the interferometer, with a relative phase determined by $\theta$. This again determines the detection probability at $\mathrm{A}_\mathrm{b}$ or $\mathrm{B}_\mathrm{b}$. A nonzero $\delta$ means that the superposition switches back and forth from symmetric to antisymmetric as $\tau$ increases, which is observable if $\delta \gtrsim \tilde{\gamma}$. Note also that the ratio $g^{(2)}_{\Ab|\Ar}/(g^{(2)}_{\Ab|\Ar} + g^{(2)}_{\Bb|\Ar})$ can exceed its classical bound of $2/3$ and come close to unity for small mean phonon numbers $n_\M$. This means that, in the limit $n_\M, \delta/\tilde{\gamma} \rightarrow 0$, one can e.g.~arrange the phases in such a way that when a red photon is detected at $\mathrm{A}_\mathrm{r}$, the next blue always arrives at $\mathrm{A}_\mathrm{b}$.

We now discuss how entanglement between the two mechanical oscillators can be detected. We wish to verify that following a red photon detection, the subsequent blue photon is in a superposition state of originating from cavity 1 and cavity 2, i.e., that there is entanglement between the output modes $\bhat_{\mathrm{b},1}$ and $\bhat_{\mathrm{b},2}$. This entanglement must be smaller than or equal to the entanglement between the mechanical oscillators. Again, we let $t' = t + \tau$. If $\tilde{\rho}(t')$ is a separable state, it is straightforward to show that
\begin{equation}
  \label{eq:Rdef}
   R(\tau) \equiv \frac{\langle \bhat^\dagger_{\mathrm{b},1}(t') \bhat_{\mathrm{b},1}(t') \bhat^\dagger_{\mathrm{b},2}(t') \bhat_{\mathrm{b},2}(t') \rangle_{\Ar}}{\big|\langle \bhat^\dagger_{\mathrm{b},1}(t') \bhat_{\mathrm{b},2}(t') \rangle_{\Ar}\big|^2}  \geq 1 \ .
\end{equation}
This also holds for classical correlations between the two fields, which e.g.~could originate from technical laser noise. Thus, we can use $R(\tau)$ as an entanglement witness \cite{Horodecki2009RMP,supplementary}.

Expectation values with respect to the state $\tilde{\rho}(t')$ can be measured through degrees of higher-order coherence, as Eq.~(\ref{eq:condonAr}) suggests. We find that $R(\tau) \leq R_\mathrm{m}(\tau)$, where the measurable upper bound for the entanglement witness is \cite{supplementary} 
\begin{equation}
   \label{eq:expValues}
   R_\mathrm{m}(\tau) = 4 \, \frac{g^{(2)}_{\Ab|\Ar}(\tau) + g^{(2)}_{\Bb|\Ar}(\tau) - 1}{\left(g^{(2)}_{\Ab|\Ar}(\tau) - g^{(2)}_{\Bb|\Ar}(\tau) \right)^2} \ ,
\end{equation}
for a symmetric setup where the output flux is the same from both cavities. Thus, a measurement of $R_\mathrm{m}(\tau) < 1$  is evidence of entanglement between the mechanical oscillators. 

\begin{figure}[htbp]
\begin{center} 
\includegraphics[width=.49\textwidth]{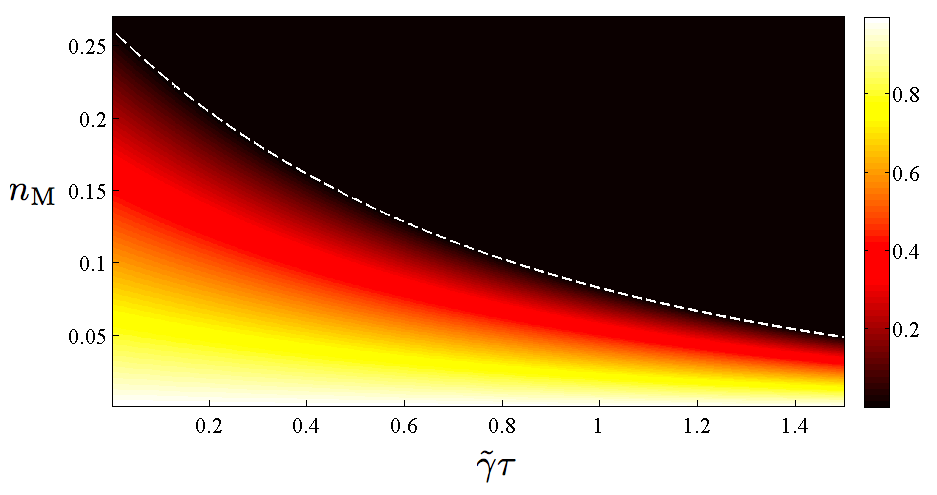}
\caption{(Color online) Density plot of $\mathrm{max}(1 - R_\mathrm{m}(\tau),0)$ as a function of mean phonon number $n_\M$ and time delay $\tau$. We set $\delta \tau + 2\phi = 0$. The separability criterion is violated below the white dashed curve, proving entanglement between the two mechanical oscillators. The entanglement can be verified for mean phonon numbers $n_\M < 0.26$ and time delays $\tau < \tilde{\gamma}^{-1} \ln ((\sqrt{2}-1)(n_\M + 1)/2n_\M)$.}
\label{fig:plot}
\end{center}
\end{figure}

We now show that the separability criterion (\ref{eq:Rdef}) is violated as the oscillators are cooled close to the ground state. Inserting the expressions in Eq.~(\ref{eq:g2Interferometer}), we arrive at
\begin{equation}
  \label{eq:Rcalculated}
  R_\mathrm{m}(\tau) =  \frac{ 4 n_\M \left[n_\M + (n_\M + 1)\ex^{-\tilde{\gamma} \tau} \right]}{ (n_\M + 1)^2\ex^{-2\tilde{\gamma} \tau} \cos^2(\delta \tau + 2\phi)}  \ ,
\end{equation}
again for $\tau \gg \kappa^{-1}, \lambda^{-1}$. In Fig.~\ref{fig:plot} we plot $\mathrm{max}(1 - R_\mathrm{m}(\tau),0)$ as a function of phonon number $n_\M$ and delay time $\tau$ in units of $\tilde{\gamma}^{-1}$, when assuming $\delta \tau + 2\phi$ to be an integer of $\pi$. We observe that entanglement can be verified through violation of the separability criterion for mean phonon numbers $n_\M < 0.26$. For $n_\M \ll 1$, the entanglement can be detected for times $\tau < \tilde{\gamma}^{-1} \ln ((\sqrt{2}-1)/2n_\M)$. 
It is also worth mentioning that before the first blue photon is emitted, the entanglement between the mechanical oscillators only decays on the timescale of the mechanical phase decoherence, which is likely to be much larger than $\tilde{\gamma}^{-1}$. This could be measurable by looking at the statistics of only the first blue photon following a red.

To show that our proposed experiment is within reach of present day experiments, we take the membrane-in-the-middle geometry \cite{Thompson2008Nature} as an example. A set of realistic parameters is $\omega_\M/2\pi = 2$ MHz, $\omega_\M/\gamma = 2 \cdot 10^7$, $\kappa/2\pi = 1$ MHz, and $|\alpha|/2\pi = 10$ kHz. Assuming an initial temperature of 20 mK, this gives an effective mean phonon number $n_\M = 0.068$, with $n_\mathrm{opt} = 0.016$. The output flux of red (blue) photons would be $41$ ($172$) photons per second. The separability criterion would be violated for times $\tau < 0.47$ milliseconds. Note that this is over 400 times longer than the entanglement lifetimes reported in a corresponding experiment with atomic ensembles by Chou {\it et al.} \cite{Chou2005Nature}. The long decoherence time is a result of high mechanical quality factors even when laser cooling to the quantum regime. 

In conclusion, we have proposed an experiment for entangling remote mechanical oscillators. This would be an important milestone in the endeavour to explore quantum effects in macroscopic systems. The entanglement is induced by optical measurements and can be verified through second-order coherences of the optical field. Our proposal is relevant to present day experimental setups. We estimate entanglement storage lifetimes of milliseconds for the membrane-in-the-middle geometry, which could be of technological interest.

We thank Jack Harris, Jack Sankey and Eran Ginossar for valuable input and acknowledge support from the Research Council of Norway, FRINAT Grant 191576/V30 (KB), and from NSF, Grants DMR-1004406, 0653377 (AN,SMG). 

%\bibliographystyle{abbrv}
%\bibliography{../../interferometer}

% \newpage 
% \title[Supplementary material to ``Proposal for entangling remote micromechanical oscillators via optical measurements'']
% {Supplementary material to ``Proposal for entangling remote\\ micromechanical oscillators via optical measurements''}
% \author{K. B{\o}rkje}
% \author{A. Nunnenkamp}
% \author{S. M. Girvin}
% \affiliation{Departments of Physics and Applied Physics, Yale University, New Haven, Connecticut 06520, USA}

% \maketitle
\newpage
\begin{center}
{\large {\bf Supplementary Material to ``Proposal for entangling remote micromechanical\\ oscillators via optical measurements''}}\\
\end{center}

\section{Introduction}

This is a supplementary to the Letter ``Proposal for entangling remote micromechanical oscillators via optical measurements''. In Section \ref{sec:resolved}, we briefly point out why our proposed experiment can also be relevant for systems that are not in the resolved sideband regime. Section \ref{sec:details} provides some details on how to calculate the correlation functions presented in the Letter. In Section \ref{sec:reqFilter}, the challenge of filtering out the sidebands into separate spatial modes is addressed quantitatively, whereas Section \ref{sec:exFilter} presents a specific example of such a filter. Most importantly, in Section \ref{sec:heterodyne} we show how the degrees of second-order coherence discussed in the Letter can be obtained in heterodyne photodetection without the need for a physical frequency filter. Finally, Section \ref{sec:entanglement} presents details on how our separability criterion can be related to observable correlation functions, and how it connects to a more familiar entanglement measure, namely concurrence.

\section{Resolved vs unresolved sideband regime}
\label{sec:resolved}

In the Letter, we focus on the situation where the temperature of the bath is large compared to the mechanical frequency, such that $n_\mathrm{th} \gg 1$. The oscillator then needs to be laser cooled by detuning the laser to $\Delta = -\omega_\M$. We consider the regime where the optical contribution to the mechanical damping $\gamma_\mathrm{opt}$ is large compared to the intrinsic damping rate $\gamma$ to the thermal bath. In that limit, the effective phonon number is
\begin{equation}
  \label{eq:nMResolved}
  n_\M \approx \frac{\gamma}{\gamma_\mathrm{opt}} n_\mathrm{th} + n_\mathrm{opt}  \ .
\end{equation}
To reach the quantum regime $n_\M < 1$, one must then be in the resolved sideband regime $\omega_\M > \kappa$, such that $n_\mathrm{opt} = \left(\kappa/4\omega_\M\right)^2$ is small.

However, one can also imagine a scenario where $n_\mathrm{opt} < 1$ is not necessary to obtain entanglement. The oscillator would then need to be coupled to a very cold thermal bath, such that $n_\mathrm{th} \ll 1$. This is particularly relevant to optomechanical realizations with large mechanical frequencies. If one then operates in a parameter regime where the intrinsic damping rate $\gamma$ is large compared to the optically induced damping rate $\gamma_\mathrm{opt}$ (e.g.~through weak driving), the effective phonon number would be
\begin{equation}
  \label{eq:nMNonResolved}
  n_\M \approx n_\mathrm{th} + \frac{\gamma_\mathrm{opt}}{\gamma} n_\mathrm{opt}  \ ,
\end{equation}
which could still be small enough to achieve entanglement.

\section{Details of the calculation}
\label{sec:details}
In this section, we present some details on how to calculate the correlation functions discussed in the Letter.

\subsection{Single cavity}
We first consider the case of a single cavity. The output field at the right-hand side of the cavity in a frame rotating at the drive frequency $\omega_\D$ is $\tilde{b}(t) = \ex^{\im \omega_\D t} \bhat(t)$. Its Fourier transform is defined as
\begin{equation}
  \label{eq:bFourierDef}
  \bhat[\omega] = \int_{-\infty}^{\infty} \du t \, \ex^{\im \omega t} \tilde{b}(t) \ .
\end{equation}
We also define
\begin{equation}
  \label{eq:bFourierDefAdj}
  \bhat^\dagger[\omega] = \int_{-\infty}^{\infty} \du t \, \ex^{\im \omega t} \tilde{b}^\dagger(t) \ , 
\end{equation}
and the spectra
\begin{equation}
  \label{eq:Sbbdefff}
  S_{\tilde{b}^\dagger \tilde{b}}[\omega] = \int_{-\infty}^{\infty} \du t \, \ex^{\im \omega t} \langle \tilde{b}^\dagger(t) \tilde{b}(0) \rangle 
\end{equation}
and 
\begin{equation}
  \label{eq:Sbbdefff2}
  S_{\tilde{b}^\dagger \tilde{b}^\dagger}[\omega] = \int_{-\infty}^{\infty} \du t \, \ex^{\im \omega t} \langle \tilde{b}^\dagger(t) \tilde{b}^\dagger(0) \rangle \ .
\end{equation}
These spectra can be found by solving the quantum Langevin equations in Fourier space and using the properties of the noise operators $\hat{c}_\mathrm{in}$ and $\xihat_i$. Let us define the bare cavity susceptibility as
\begin{equation}
  \label{eq:chiC}
  \chi_\C[\omega] = \frac{1}{\kappa/2 - \im (\omega + \Delta)} \ ,
\end{equation}
where $\Delta = \omega_\D - \omega_\C - g\langle \hat{x} \rangle$ is the detuning. We also define the effective mechanical susceptibility
\begin{equation}
  \label{eq:chiM}
  \tilde{\chi}_\M[\omega] = \frac{1}{\tilde{\gamma}/2 - \im (\omega - \tilde{\omega}_\M)} \ .
\end{equation}
For the spectrum (\ref{eq:Sbbdefff}), we have $S_{\tilde{b}^\dagger \tilde{b}}[\omega] = \kappa_\RR |\bar{a}|^2 2 \pi \delta(\omega) + \kappa_\RR |\alpha|^2 x_\mathrm{zpf}^{-2} |\chi_\C[-\omega]|^2 S_{\hat{x}\hat{x}}[\omega]$, where the mechanical oscillator spectral density is approximately
\begin{equation}
  \label{eq:oscS}
  S_{\hat{x}\hat{x}}[\omega] = x_\mathrm{zpf}^{2} \tilde{\gamma} \Big[|\tilde{\chi}_\M[\omega]|^2 \left( n_\M + 1 \right) + |\tilde{\chi}[-\omega]|^2 n_\M \Big]
\end{equation}
in the weak coupling limit. 

The output filtered around the blue (red) sideband is defined by $\bhat_\mathrm{b(r)}(t) = \ex^{-\im \omega_\D t} \tilde{b}_\mathrm{b(r)}(t)$ with
\begin{equation}
  \label{eq:bbDef}
  \tilde{b}_\mathrm{b}(t) = \frac{1}{2 \pi} \int_{-\infty}^{\infty} \du \omega \, \ex^{-\im \omega t} F_\mathrm{b}[\omega] \bhat[\omega] 
\end{equation}
and
\begin{equation}
  \label{eq:brDef}
  \tilde{b}_\mathrm{r}(t) = \frac{1}{2 \pi} \int_{-\infty}^{\infty} \du \omega \, \ex^{-\im \omega t} F_\mathrm{r}[\omega] \bhat[\omega] \ .
\end{equation}
The filter functions $F_\mathrm{b(r)}[\omega]$ will naturally depend on the specific details of the filter. They should have the properties that $F_\mathrm{b(r)}[\omega] \sim 1$ for frequencies $\omega$ in a range of size $> \tilde{\gamma}$ around $\pm \omega_\M$. On the other hand, $|F_\mathrm{b(r)}[\omega]|$ should be as small as possible for $\omega \sim 0$, such that the carrier photons at the drive frequency $\omega_\D$ are rejected. The quantitative requirements for these filter functions are discussed in Section \ref{sec:reqFilter}. In the following, we will assume that the filters are good enough to only pick out the sidebands, such that we can neglect the carrier entirely.

Since we linearize the equations of motion and only consider Gaussian input, Wick's theorem allows all higher-order correlation functions to be written in terms of correlation functions containing two operators. As an example, consider 
\begin{eqnarray}
  \label{eq:g2ex}
  g^{(2)}_\mathrm{b|r}(\tau) & = & \frac{\langle \bhat^\dagger_\rr(t) \bhat^\dagger_\bb(t + \tau) \bhat_\bb(t + \tau) \bhat_\rr(t)\rangle}{\langle \bhat^\dagger_\rr(t) \bhat_\rr(t)\rangle \langle \bhat^\dagger_\bb(t) \bhat_\bb(t)\rangle} \\
  & = & 1 + \frac{|\langle \tilde{b}^\dagger_\rr(t) \tilde{b}_\bb(t + \tau) \rangle|^2 + |\langle \tilde{b}^\dagger_\rr(t) \tilde{b}^\dagger_\bb(t + \tau) \rangle|^2}{\langle \tilde{b}^\dagger_\rr(t) \tilde{b}_\rr(t)\rangle \langle \tilde{b}^\dagger_\bb(t) \tilde{b}_\bb(t)\rangle} \notag \ .
\end{eqnarray}
We will repeatedly make use of the fact that the effective mechanical linewidth $\tilde{\gamma}$ is much smaller than the cavity linewidth $\kappa$, as well as the integral
\begin{equation}
  \label{eq:Integral}
 \int_{-\infty}^{\infty} \frac{\du \omega}{2\pi}  \, \ex^{\im \omega \tau}  \tilde{\gamma} |\tilde{\chi}_\M[\pm\omega]|^2 = \ex^{-(\tilde{\gamma}/2 \mp \im \tilde{\omega}_\M)\tau} \ ,
\end{equation}
valid for $\tau \geq 0$. We start with the red photon flux, which is given by %\begin{widetext}
\begin{eqnarray}
  \label{eq:brDbr}
  \langle \tilde{b}^\dagger_\rr(t) \tilde{b}_\rr(t) \rangle & = & \int_{-\infty}^{\infty} \frac{\du \omega}{2\pi} \, |F_\rr[-\omega]|^2 S_{\tilde{b}^\dagger \tilde{b}}[\omega] \\ 
   & = & \kappa_\RR |\alpha|^2 |\chi_\C[-\omega_\M]|^2 (n_\M + 1) \notag \ . 
\end{eqnarray}
Similarly, the blue photon flux becomes
\begin{eqnarray}
  \label{eq:bbDbb}
\langle \tilde{b}^\dagger_\bb(t) \tilde{b}_\bb(t) \rangle & = & \int_{-\infty}^{\infty} \frac{\du \omega}{2\pi} \, |F_\bb[-\omega]|^2 S_{\tilde{b}^\dagger \tilde{b}}[\omega] \\
 & = & \kappa_\RR |\alpha|^2 |\chi_\C[\omega_\M]|^2 \, n_\M \notag \ . 
\end{eqnarray}
Moving on to the cross-correlators, we find
\begin{eqnarray}
  \label{eq:brDbbD}
  & & \langle \tilde{b}^\dagger_\rr(t) \tilde{b}^\dagger_\bb(t + \tau) \rangle =  \int_{-\infty}^{\infty}  \frac{\du  \omega}{2\pi} \, \ex^{\im \omega \tau} F_\rr^\ast[-\omega] F_\bb^\ast[\omega] S_{\tilde{b}^\dagger \tilde{b}^\dagger}[\omega] \quad  \notag \\
  & & \quad \approx  - \kappa_\RR \alpha^{\ast \, 2} \int_{-\infty}^{\infty} \frac{\du \omega}{2\pi} \, \ex^{\im \omega \tau} \chi_\rr^\ast[-\omega] \chi_\bb^\ast[\omega] \chi^\ast_\C[\omega] \chi^\ast_\C[-\omega] \notag \\
 &  & \qquad \qquad \qquad  \times  \Big[\tilde{\gamma} \, |\tilde{\chi}_\M[\omega]|^2  (n_\M + 1) - \tilde{\chi}_\M[\omega] \Big] 
\end{eqnarray} 
We assume that the delay time $\tau \gg \lambda^{-1}, \kappa^{-1}$, in which case the last term does not contribute, giving
\begin{eqnarray}
  \label{eq:brDbbD2}
 & & \langle \tilde{b}^\dagger_\rr(t) \tilde{b}^\dagger_\bb(t + \tau) \rangle \\
&  & \ \approx  - \kappa_\RR \alpha^{\ast \, 2} \ex^{-\im \vartheta} \chi^\ast_\C[\omega_\M] \chi^\ast_\C[-\omega_\M]  (n_\M + 1) \ex^{-(\tilde{\gamma} - \im \tilde{\omega_\M})\tau} \ . \notag
\end{eqnarray}
The filter functions might give rise to the phase factor $\ex^{-\im \vartheta}$, but this does not contribute in Eq.~(\ref{eq:g2ex}). Finally, we have
\begin{eqnarray}
  \label{eq:brDbb}
  \langle \tilde{b}^\dagger_\rr(t) \tilde{b}_\bb(t + \tau) \rangle & = & \int_{-\infty}^{\infty} \frac{\du \omega}{2\pi} \, \ex^{\im \omega \tau} F_\rr^\ast[-\omega] F_\bb[-\omega] S_{\tilde{b}^\dagger \tilde{b}}[\omega]  \notag \\ & \approx & 0 \ .
\end{eqnarray}
%\end{widetext} 
Inserting these results into Eq.~(\ref{eq:g2ex}) gives the expression for $g^{(2)}_{\bb|\rr}(\tau)$.

\subsection{Two cavitites}

We now move on to the two-cavity setup in Fig.~1 of the Letter. The calculation of the correlation functions $g^{(2)}_{\Ab|\Ar}(\tau)$, $g^{(2)}_{\Bb|\Ar}(\tau)$, etc. ~is performed in a similar way as above. However, one needs to relate the fields at the photodetectors to the cavity output fields via the beam splitter transfer matrix:
\begin{eqnarray}
  \label{eq:beamsplittTransfer}
  \bhat_\A(t) & = & \frac{1}{\sqrt{2}} \left(\bhat_2(t) \ex^{\im \phi} + \im \bhat_1(t) \right) \ , \\
   \bhat_\B(t) & = & \frac{1}{\sqrt{2}} \left(\bhat_1(t)  + \im \bhat_2(t) \ex^{\im \phi} \right) \ . \notag 
\end{eqnarray}
The second-order coherences $g^{(2)}_{\Ab|\Ar}(\tau)$ and $g^{(2)}_{\Bb|\Ar}(\tau)$  then follow from the correlation functions calculated in the previous subsection, as well as the fact that the fields $\bhat_1$ and $\bhat_2$ are uncorrelated in the absence of measurements (i.e.~in the state $\rho_\mathrm{ss}$).

\section{Requirements for the frequency filter}
\label{sec:reqFilter}

If one would like to detect the red and blue photons individually on separate photomultipliers, one needs to separate the optical sidebands at the frequency $\omega_\D \pm \omega_\M$ from the carrier at $\omega_\D$. This is a big technical challenge, partly because the mechanical frequencies can be in the kHz-MHz range, and, most importantly, because the carrier photons far outnumber the sideband photons in the weak coupling regime we discuss. Let us consider the photon flux of the light leaking out at the right hand side of the cavity. We denote $f_\mathrm{b}$ ($f_\mathrm{r}$) the flux of blue (red) photons and $f_\mathrm{c}$ the flux of photons at the laser frequency $\omega_\D$. 

The photon fluxes are
\begin{eqnarray}
  \label{eq:fluxes}
  f_\mathrm{b} & = & \kappa_\RR |\alpha|^2 |\chi_\C[\omega_\M]|^2 n_\M \ , \\  
  f_\mathrm{r} & = & \kappa_\RR |\alpha|^2 |\chi_\C[-\omega_\M]|^2  (n_\M + 1) \notag \ ,\\ 
  f_\mathrm{c} & = & \kappa_\RR |\bar{a}|^2 \notag  \ .
\end{eqnarray}
The cavity susceptibility $\chi_\C[\omega]$ is defined in Eq.~(\ref{eq:chiC}). Remembering that $|\alpha| = g x_\mathrm{zpf} |\bar{a}|$, this means that the ratio between blue photon flux and carrier photon flux is
\begin{equation}
  \label{eq:fluxbfluxc}
  \frac{f_\mathrm{b}}{f_\mathrm{c}} = 4 \left(\frac{g x_\mathrm{zpf}}{\kappa} \right)^2 n_\M  \ ,
\end{equation} 
for a detuning $\Delta = - \omega_\M$. The ratio between the red and blue photon flux is
\begin{equation}
  \label{eq:fluxrfluxb}
  \frac{f_\mathrm{r}}{f_\mathrm{b}} = \frac{n_\mathrm{opt} (n_\M + 1)}{(n_\mathrm{opt} + 1) n_\M} \ .
\end{equation}

According to Eq.~(\ref{eq:fluxbfluxc}), the relative flux of the sidebands to the carrier depends on the ratio between the bare optomechanical coupling $g x_\mathrm{zpf}$ and the cavity linewidth $\kappa$. We will continue with the membrane-in-the-middle setup as an example. In present-day membrane-in-the-middle experiments, the bare coupling constant $g x_\mathrm{zpf}/2\pi \sim 10 \text{ Hz}$ \cite{Harrisgroup}. If we optimistically assume that this could be increased to 100 Hz, still use $\kappa/2\pi = 1 \text{ MHz}$, and assume $n_\M = 0.1$, we find $f_\mathrm{b}/f_\mathrm{c} \sim 10^{-8}$.
% \begin{equation}
%   \label{eq:4}
%    \frac{f_\mathrm{b}}{f_\mathrm{c}} \sim 10^{-8} \ .
% \end{equation}
This indicates how challenging the filtering can be, at least in the membrane-in-the-middle setup.

\section{The frequency filter - an example}
\label{sec:exFilter}
In this section, we present an example of a specific filter that would separate the sidebands into different spatial modes. We do not claim that this is the best way to perform the sideband filtering. However, we wish to argue that, despite the considerable technical challenge, the task can be within reach of experiments. 

\begin{figure}[htbp]
\begin{center} 
\includegraphics[width=.42\textwidth]{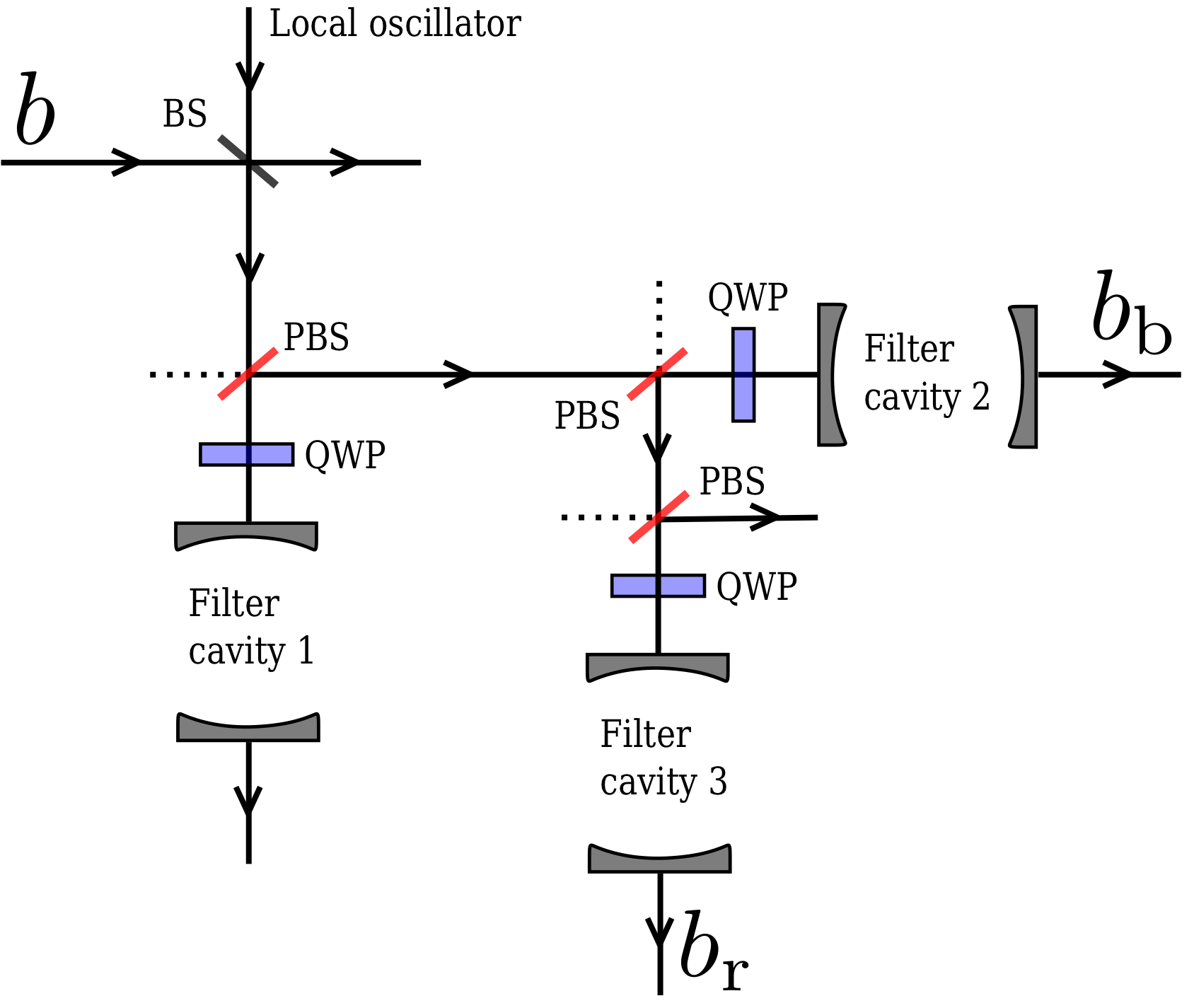}
\caption{(Color online) A simple suggestion for how to filter out the sidebands from the carrier. The signal is combined with a local oscillator on a beam splitter (BS) with high reflectivity to null out the carrier as much as possible through destructive interference. It then proceeds to filter cavity 1, which mostly transmits carrier photons while reflecting the red and blue sidebands. The quarter wave plate (QWP) and the polarizing beam splitter (PBS) ensure that none of the reflected photons are lost. Filter cavity 2 transmits the blue sideband, whereas cavity 3 transmits the red sideband.}
\label{fig:filter}
\end{center}
\end{figure}

We imagine a filter consisting of four stages, as depicted in Fig.~\ref{fig:filter}.
The first stage is to combine the output field $\bhat(t)$ with a local oscillator (split off from the original laser) on a beam splitter with very high reflectivity. By adjusting the power and phase of the local oscillator, one can remove a significant part of the carrier at this stage. Ref.~\cite{Wittmann2010PRL} reported an extinction ratio of 1/700 by this method. We will assume that the carrier flux is reduced by the factor $R$ by this displacement.

The next stage of our filter is the optical cavity labeled ``Filter cavity 1'' in Fig.~\ref{fig:filter}. The purpose of this cavity is to have the sidebands be promptly reflected, whereas a large part of the carrier is transmitted through the cavity. Filter cavity 1 has a mode at the frequency $\omega_1$ with the linewidth $\mu$. We require $\Gamma_\D \ll \mu \ll \omega_\M$ with $\Gamma_\D$ being the laser linewidth. Also, we let $\omega_1 = \omega_\D - \Delta_1$ with $\Delta_1 \ll \mu$. Finally, we denote the decay rate through the upper mirror $\mu_\mathrm{U}$ and assume that it is $\mu_\mathrm{U} = \mu(1/2 + \delta_1)$, i.e.~approximately half the total cavity decay rate, with a deviation $\delta_1 \ll 1$. We assume that the initial signal $\bhat(t)$ is polarized. The polarizing beam splitter (PBS) and the quarter-wave plate (QWP) then ensure that whatever is reflected from the cavity is also reflected by the PBS afterwards, such that we (in the ideal case) do not lose any of the photons that are reflected. We neglect complications that might arise from a polarization mismatch between the incoming field and the internal cavity field. The vacuum noise at the PBS can be neglected, since it does not produce photodetection events. 

The purpose of filter cavity 2 is to let the blue sideband be transmitted, but reflect the red sideband plus what is left of the carrier. Filter cavity 2 has a mode at the frequency $\omega_2 = \omega_\D + \omega_\M$ and a linewidth $\lambda$ with the requirement $\tilde{\gamma} \ll \lambda \ll \omega_\M$. Ideally, it should also be symmetric and have minimal loss, such that the left and right decay rates are equal, i.e.~$\lambda_\LL \approx \lambda_\RR$, and $ \lambda_\LL = \lambda(1/2 + \delta_2)$ with $\delta_2 \ll 1$. Filter cavity 3 has the same properties as cavity 2, except that it is centered on the frequency $\omega_3 = \omega_\D - \omega_\M$, such that it transmits the red sideband.

The three filter cavities results in the filter functions (see Eqs.~(\ref{eq:bbDef}), (\ref{eq:brDef}))
\begin{eqnarray}
  \label{eq:chibchir}
  F_\mathrm{b}[\omega] & = & \frac{\sqrt{\lambda_\LL \lambda_\RR}}{\lambda/2 - \im \left(\omega - \omega_\M \right)} \rho[\omega]\\
  F_\mathrm{r}[\omega] & = &  \frac{\sqrt{\lambda_\LL \lambda_\RR}}{\lambda/2 - \im \left(\omega + \omega_\M  \right)}  \frac{\lambda \delta_2 + \im \left(\omega - \omega_\M \right)}{\lambda/2 - \im \left(\omega - \omega_\M \right)} \rho[\omega] \quad \notag
\end{eqnarray}
where we defined the function
\begin{equation}
  \label{eq:chirefl}
  \rho[\omega] = \frac{\mu \delta_1 + \im \left(\omega + \Delta_1\right)}{\mu/2 - \im \left(\omega + \Delta_1\right)} \ .
\end{equation}
After the displacement by the local oscillator, the original output spectrum in the rotating frame is 
\begin{eqnarray}
  \label{eq:spectrumOrig}
  S_{\tilde{b}^\dagger \tilde{b}}[\omega] 
  & = & \frac{\Gamma_\mathrm{D}}{\left(\Gamma_\mathrm{D}/2\right)^2 + \omega^2} \, R  f_\mathrm{c} + \frac{\tilde{\gamma}}{\left(\tilde{\gamma}/2\right)^2 + \left(\omega - \tilde{\omega}_\M\right)^2} \, f_\mathrm{r} \notag \\
  & + & \frac{\tilde{\gamma}}{\left(\tilde{\gamma}/2\right)^2 + \left(\omega + \tilde{\omega}_\M\right)^2} \, f_\mathrm{b} 
\end{eqnarray}
when we assume the laser to have a Lorentzian lineshape with a width $\Gamma_\mathrm{D}$. After the filtering, the blue output spectrum becomes
\begin{equation}
  \label{eq:spectrumBlue}
  S_{\tilde{b}_\mathrm{b}^\dagger \tilde{b}_\mathrm{b}}[\omega] =  |\chi_\mathrm{b}[-\omega]|^2 S_{\tilde{b}^\dagger \tilde{b}}[\omega] \ .
\end{equation}
By integrating over the (suppressed) carrier peak, we get the flux of remaining carrier photons in the blue output,
\begin{eqnarray}
  \label{eq:fluxcB}
  f^\mathrm{(b)}_\mathrm{c} & \equiv & \frac{1}{2\pi} \int_\mathrm{c} \du \omega \, S_{\tilde{b}_\mathrm{b}^\dagger \tilde{b}_\mathrm{b}}[\omega] \\
   & \approx & \frac{\lambda_\LL \lambda_\RR}{\omega_\M^2} \left(\frac{\Gamma_\mathrm{D}}{\mu} + 4 \delta_1^2\right) R f_\mathrm{c} \ . \notag
\end{eqnarray}
Here, the subscript c on the integration sign denotes that only the carrier peak is integrated over. We have also assumed that $\Gamma_\mathrm{D}, \lambda \ll \omega_\M$. With these approximations, we reach the same expression for $f^\mathrm{(r)}_\mathrm{c}$, the remaining flux of carrier photons in the red output. 

Let us use the same numerical example as in the Letter, where $\omega_\M/2\pi = 2 \text{ MHz}$ and $\tilde{\gamma}/2\pi \approx 0.4 \text{ kHz}$. We assume that the filter cavities have widths $\mu/2\pi = \lambda/2\pi = 10 \text{ kHz}$. Furthermore, we assume that the laser linewidth is $\Gamma_\mathrm{D} = 0.1 \text{ kHz}$ and that $4\delta_1^2 < 10^{-2}$. If we also assume that the initial displacement reduces the carrier photon flux by $R = 10^{-3}$, we arrive at $f^\mathrm{(b)}_\mathrm{c}/f_\mathrm{c} \sim 10^{-10}$, giving $f_\mathrm{b}/f^\mathrm{(b)}_\mathrm{c} \sim 10^2$. This shows that the carrier can in principle be removed from the sidebands in this idealized setup. It should nevertheless be mentioned that total suppression of reflection from a cavity is in practice quite difficult, and a minimal reflection of about 10\% is typical \cite{Harrisgroup}.

\section{Heterodyne photodetection}
\label{sec:heterodyne}

In this section, we point out that the degrees of second-order coherence discussed in the Letter are accessible through heterodyne photodetection, where the blue and red sidebands can easily be distinguished in the Fourier domain with a digital filter function of our choice. This means that the measurement-induced entanglement can be detected also in cases where filtering of the sidebands into separate spatial modes proves to be too difficult. 

\subsection{Single detector}

First, let us discuss the case of a single cavity. Assume that the output from the cavity is combined with a strong local oscillator on a beam splitter with reflectivity $r$. The transmitted part of the local oscillator and the reflected part of the signal beam are then detected at a photomultiplier. We let the local oscillator frequency be $\omega_\mathrm{LO} = \omega_\D - \omega_\mathrm{IF}$, detuned from the original drive frequency by the intermediate frequency $\omega_\mathrm{IF}$. This mixes the carrier frequency down to $\omega_\mathrm{IF}$, and the sidebands to $\omega_\mathrm{IF} \pm \omega_\M$. We will denote the photocurrent operator by $i(t)$ and its Fourier transform by $i[\omega]$. From this, we can define 
\begin{eqnarray}
  \label{eq:redblueCurrents}
  i_{\mathrm{r}}(t)  & = & \int_{-\infty}^{\infty} \frac{\du \omega}{2\pi}  \, \ex^{-\im \omega t} F_\rr[\omega] i[\omega_\mathrm{IF} + \omega] \\
  i_{\mathrm{b}}(t)  & = & \int_{-\infty}^{\infty}  \frac{\du \omega}{2\pi} \, \ex^{-\im \omega t} F_\bb[\omega]  i[\omega_\mathrm{IF} + \omega]   \ .
\end{eqnarray}
where the filter functions $F_\rr[\omega]$ and $F_\bb[\omega]$ can be chosen freely. As before, we should make sure that $F_{\bb(\rr)}[\omega] \sim 1$ over the width of the sidebands at $\omega = \pm \omega_\M$, and that they have minimal support at $\omega \sim 0$. More specifically, it is necessary that $|F_\bb[0]| \ll (f_\bb/f_\mathrm{c})^{(1/2)}$, where $f_\bb$ $(f_\mathrm{c})$ is the flux of blue (carrier) photons, and similarly for the red filter function. 

Note that $i_{\mathrm{r}}(t)$ and $i_{\mathrm{b}}(t)$ are not directly measurable quantitites, but are complex functions that can be calculated from measurements of the photocurrent $i(t)$. We can also express these ``sideband currents'' as
\begin{equation}
  \label{eq:redblueCurrentsTime}
  i_{j}(t) = \int_{-\infty}^\infty \du t_1 \, F_{j}(t-t_1) \ex^{\im \omega_\mathrm{IF} t_1} i(t_1) \ , \quad j=\rr,\bb
\end{equation}
where the filter functions in the time domain are
\begin{equation}
  \label{eq:ffTime}
  F_{j}(t) = \int_{-\infty}^\infty \frac{\du \omega}{2 \pi} \, \ex^{-\im \omega t} F_j[\omega] \ .
\end{equation}
The filter functions $F_\bb$ and $F_\rr$ should be chosen such that they are sufficiently localized both in frequency and time, meaning that $F_j[\omega]$ falls off quickly compared to the mechanical frequency $\omega_\M$, whereas $F_j(t)$ falls off quickly compared to the mechanical lifetime $1/\tilde{\gamma}$.  A suitable choice would be Gaussian functions,
\begin{equation}
  \label{eq:Gaussian}
  F_{\bb(\rr)}[\omega] = \ex^{-(\omega \mp \omega_\M)^2/(2 \lambda^2)} \ , 
\end{equation}
with the width satisfying $\tilde{\gamma} \ll \lambda \ll \omega_\M$. With the numbers used above ($\omega_\M/2\pi = 2$ MHz, $\lambda/2\pi = 10$ kHz), the filter functions at the drive frequency become $F_{\bb(\rr)}[0] = \mathrm{exp}(-2 \cdot 10^4)$, which would fully suppress the carrier. In the time domain, the filter functions are
\begin{equation}
  \label{eq:GaussianTime}
  F_{\bb(\rr)}(t) = \frac{\lambda}{2 \pi} \ex^{\mp \im \omega_\M t - \lambda^2 t^2/2} \ .
\end{equation}
which falls off fast enough that one can resolve photons on timescales much smaller than $1/\tilde{\gamma}$.

We start by calculating the expectation value
\begin{eqnarray}
  \label{eq:irSq}
  & & \langle  i_{\rr}^\dagger(t)  i_{\rr}(t)  \rangle = \int_{-\infty}^{\infty} \du t_1 \int_{-\infty}^{\infty}  \du t_2 \, \ex^{- \im \omega_\mathrm{IF} (t_1 - t_2)} \\ 
  & & \qquad \qquad \qquad \times F^\ast_{\rr}(t-t_1) F_{\rr}(t - t_2) \langle i(t_1) i(t_2) \rangle \ . \notag 
\end{eqnarray}
By the same approach as in Ref.~\cite{Carmichael1987JOptSocAmB}, one can show that the photocurrent autocorrelation function is 
\begin{eqnarray}
  \label{eq:i1i2}
  & & \langle i(t_1) i(t_2) \rangle  \\
  &  & \quad = (Ge)^2 \eta \Big(\eta \langle : I(t_1) I(t_2) : \rangle + \langle : I(t_1) : \rangle \delta (t_1 - t_2) \Big) \ . \notag
\end{eqnarray}
Here, $G$ is the photodetector gain, $e$ is the electron charge, and $\eta$ is the dimensionless photodetection efficiency. The colons indicate time and normal ordering \cite{Carmichael1987JOptSocAmB}. The operator $I(t)$ describes the photon flux at the photodetector at time $t$ and is given by
\begin{eqnarray}
  \label{eq:Idef}
  & & I(t) = (1-r) |\beta_\mathrm{lo}|^2 \\ &  & + \ \im \sqrt{r(1-r)} |\beta_\mathrm{lo}| \left(\ex^{-\im (\omega_\mathrm{IF} t + \theta_\mathrm{lo})} \tilde{b}(t) - \ex^{\im(\omega_\mathrm{IF} t + \theta_\mathrm{lo})} \tilde{b}^\dagger(t) \right) \notag
\end{eqnarray}
in the limit where the (classical) local oscillator field transmitted by the beamsplitter, $\sqrt{1-r} |\beta_\mathrm{lo}| \ex^{\im \theta_\mathrm{lo}}$, is strong compared to the output field from the cavity. The operator $\tilde{b}(t)$ is the cavity output field operator in the rotating frame. The phase $\theta_\mathrm{lo}$ is dependent on the path-length difference between the local oscillator and the signal beam, but is not important for our purpose. The integrand in Eq.~(\ref{eq:irSq}) has several terms that oscillate at the frequency $\omega_\mathrm{IF}$. In the limit $\omega_\mathrm{IF} \gg \omega_\M$, these terms can be neglected and one arrives at
\begin{eqnarray}
  \label{eq:Lambdarr}
\langle  i_{\rr}^\dagger(t)  i_{\rr}(t)  \rangle & = &  (Ge)^2 \eta t |\beta_\mathrm{lo}|^2 \left( \frac{\lambda}{2 \pi} + r \eta \langle \tilde{b}^\dagger_\rr(t) \tilde{b}_\rr(t) \rangle \right) \ , \qquad
\end{eqnarray}
where $t = 1-r$ and
\begin{equation}
  \label{eq:brTildedef}
  \tilde{b}_\rr(t) = \int_{-\infty}^\infty \du t_1 \, F_\rr(t - t_1) \tilde{b}(t_1)
\end{equation}
is the time domain version of Eq.~(\ref{eq:brDef}).

The last term in Eq.~(\ref{eq:Lambdarr}) is proportional to the flux of red sideband photons leaking out of the cavity, whereas the first term is a shot noise term. In practice, the way to get rid of the shot noise term is to integrate the photocurrent spectrum $S[\omega]$ around the frequency $\omega_\rr = \omega_\mathrm{IF} - \omega_\M$ after subtracting the shot noise floor. We denote this integrated sideband weight by $W_{\rr}$ and the shot noise floor by $S_0 \equiv (G e )^2 \eta t |\beta_\mathrm{lo}|^2$. We then find that 
\begin{equation}
  \label{eq:brbrfromSpectrum}
   W_{\rr}  \equiv \int_{\omega_\rr - \lambda/2}^{\omega_\rr + \lambda/2} \du \omega \, \left(S[\omega] - S_0 \right) =  r \eta S_0 \langle \tilde{b}^\dagger_\rr(t) \tilde{b}_\rr(t) \rangle \ .
\end{equation}
In exactly the same way, one can show that
\begin{eqnarray}
  \label{eq:Lambdabb}
  & & \langle i^\dagger_{\bb}(t) i_{\bb}(t)  \rangle =  S_0 \left( \frac{\lambda}{2 \pi} + r \eta \langle \tilde{b}^\dagger_\bb(t) \tilde{b}_\bb(t) \rangle \right) \ . 
\end{eqnarray}
with the blue sideband operator given by 
\begin{equation}
  \label{eq:bbTildedef}
  \tilde{b}_\bb(t) = \int_{-\infty}^\infty \du t_1 \, F_\bb(t - t_1) \tilde{b}(t_1) \ .
\end{equation}
The term containing the blue photon flux can also be determined from the photocurrent spectrum by
\begin{equation}
  \label{eq:bbbbfromSpectrum}
  W_{\bb } \equiv \int_{\omega_\bb - \lambda/2}^{\omega_\bb + \lambda/2} \du \omega \, \left(S[\omega] - S_0 \right) =  r \eta S_0 \langle \tilde{b}^\dagger_\bb(t) \tilde{b}_\bb(t) \rangle \ ,
\end{equation}
where we have defined $\omega_\bb = \omega_\mathrm{IF} + \omega_\M$.

Next, we define the correlation functions
\begin{eqnarray}
  \label{eq:1}
  \Lambda_{jk} (\tau) & = & \langle i^\dagger_{j}(t) i_{k}(t + \tau) \rangle  \ , \\
  \Gamma_{jk} (\tau) & = & \langle i^\dagger_{j}(t) i^\dagger_{k}(t + \tau) \rangle \ , \notag
\end{eqnarray}
where $j,k = \rr,\bb$. These functions can be determined from measurements of the photocurrent $i(t)$. We will assume that the delay time $\tau \gg \lambda^{-1}$, such that $F_j(\tau) \approx 0$. In that case, one can show that %\begin{widetext}
\begin{eqnarray}
  \label{eq:LambdarrTau}
  \Lambda_{jk} (\tau) & = & r \eta S_0 \langle \tilde{b}^\dagger_j(t) \tilde{b}_k(t + \tau) \rangle   \ , \\
  \Gamma_{jk} (\tau) & = & - r \eta S_0  \ex^{2\im \theta_\mathrm{lo}} \langle \tilde{b}^\dagger_j(t) \tilde{b}^\dagger_k(t + \tau) \rangle  \ . \notag
\end{eqnarray}

These correlation functions are all we need to determine the degrees of second-order coherence $g^{(2)}_{j|i}(\tau)$, with $i,j = \rr,\bb$ and $\tau \gg \lambda^{-1}$, whenever $\rho_\mathrm{ss}$ is a Gaussian state. Specifically, we have
\begin{eqnarray}
  \label{eq:g2byLambdaGamma}
  g^{(2)}_{j|i}(\tau) = 1 + \frac{|\Lambda_{ij}(\tau)|^2 + |\Gamma_{ij}(\tau)|^2}{W_{i} W_{j}} \ .
\end{eqnarray}
Notice that it is not necessary to determine the parameters $G$, $r$ and $\eta$. Furthermore, the second-order coherences are not susceptible to a frequency-dependent gain, as long as the gain is approximately constant over a frequency window of width $\tilde{\gamma}$.

\subsection{Multiple detectors}

In the two-cavity setup of the Letter, we are also interested in observing cross-correlators such as $g^{(2)}_{\Bb|\Ar}(\tau)$. In that case, one has to perform heterodyne detection on both the fields $\bhat_\mathrm{A}(t)$ and $\bhat_\mathrm{B}(t)$, which are the cavity output fields after having been combined at the beam splitter (see Fig.~1 of the Letter and Eq.~(\ref{eq:beamsplittTransfer}) of Section \ref{sec:details}). We assume that this is carried out in the way discussed above, giving two photocurrents $i_\A(t)$ and 
$i_\B(t)$, which have spectra $S_\A[\omega]$ and $S_\B[\omega]$, respectively. 

The integrated red sideband weight of the spectrum $S_\A[\omega]$ is 
\begin{eqnarray}
  \label{eq:LambdaArAr}
  W_{\Ar} & \equiv & \int_{\omega_\rr - \lambda/2}^{\omega_\rr + \lambda/2} \du \omega \, \left(S_\A[\omega] - S_0 \right) \\ & = &  r \eta S_0  \langle \tilde{b}^\dagger_{\Ar}(t) \tilde{b}_{\Ar}(t) \rangle\ , \notag
\end{eqnarray}
and similarly for the blue sideband as well as the sidebands of the spectrum $S_\B[\omega]$. As above, we can define correlation functions from the filtered ``sideband currents'' $i^\A_{\rr}$, $i^\A_{\bb}$, $i^\B_{\rr}$, and $i^\B_{\bb}$:
\begin{eqnarray}
  \label{eq:1}
  \Lambda_{\mathrm{A}_j \mathrm{B}_k} (\tau) & \equiv & \langle i^{\A \, \dagger}_{j}(t) i^\B_{k}(t + \tau) \rangle   \\
     & = & r \eta S_0 \langle \tilde{b}^\dagger_{\A_j}(t) \tilde{b}_{\B_k}(t + \tau) \rangle  \ ,  \notag \\ 
  \Gamma_{\mathrm{A}_j \mathrm{B}_k} (\tau) & = & \langle i^{\A \, \dagger}_{j}(t) i^{\B \, \dagger}_{k}(t + \tau) \rangle \notag \\
  & = & - r \eta S_0  \ex^{\im (\theta_{\mathrm{lo},\A} + \theta_{\mathrm{lo},\B} )} \langle \tilde{b}^\dagger_{\A_j}(t) \tilde{b}^\dagger_{\B_k}(t + \tau) \rangle  \ . \notag
\end{eqnarray}
From these correlation functions, the degrees of second-order coherence can be found. As an example, we have
\begin{equation}
  \label{eq:g2ABHD}
  g^{(2)}_{\Bb|\Ar}(\tau) = 1 + \frac{|\Lambda_{\Ar\Bb}(\tau)|^2 + |\Gamma_{\Ar\Bb}(\tau)|^2}{W_{\Ar} W_{\Bb}} \ .
\end{equation}
Although we have for simplicity assumed that the local oscillator strength is the same for both the A and B fields, it should be noted that the second-order coherences are not susceptible to differences in these.

\section{Entanglement witness}
\label{sec:entanglement}
In this section, we discuss the entanglement witness $R(\tau)$, defined in Eq.~(6) of the Letter. In Section \ref{sec:blabla}, we show how it can be expressed in terms of measurable higher-order coherences of the electromagnetic field. Section \ref{sec:blablabla} relates our separability criterion to an entanglement measure on the state of the two mechanical oscillators.

\subsection{Relation to degrees of higher-order coherence}
\label{sec:blabla}
\begin{figure}[htbp]
\begin{center} 
\includegraphics[width=.47\textwidth]{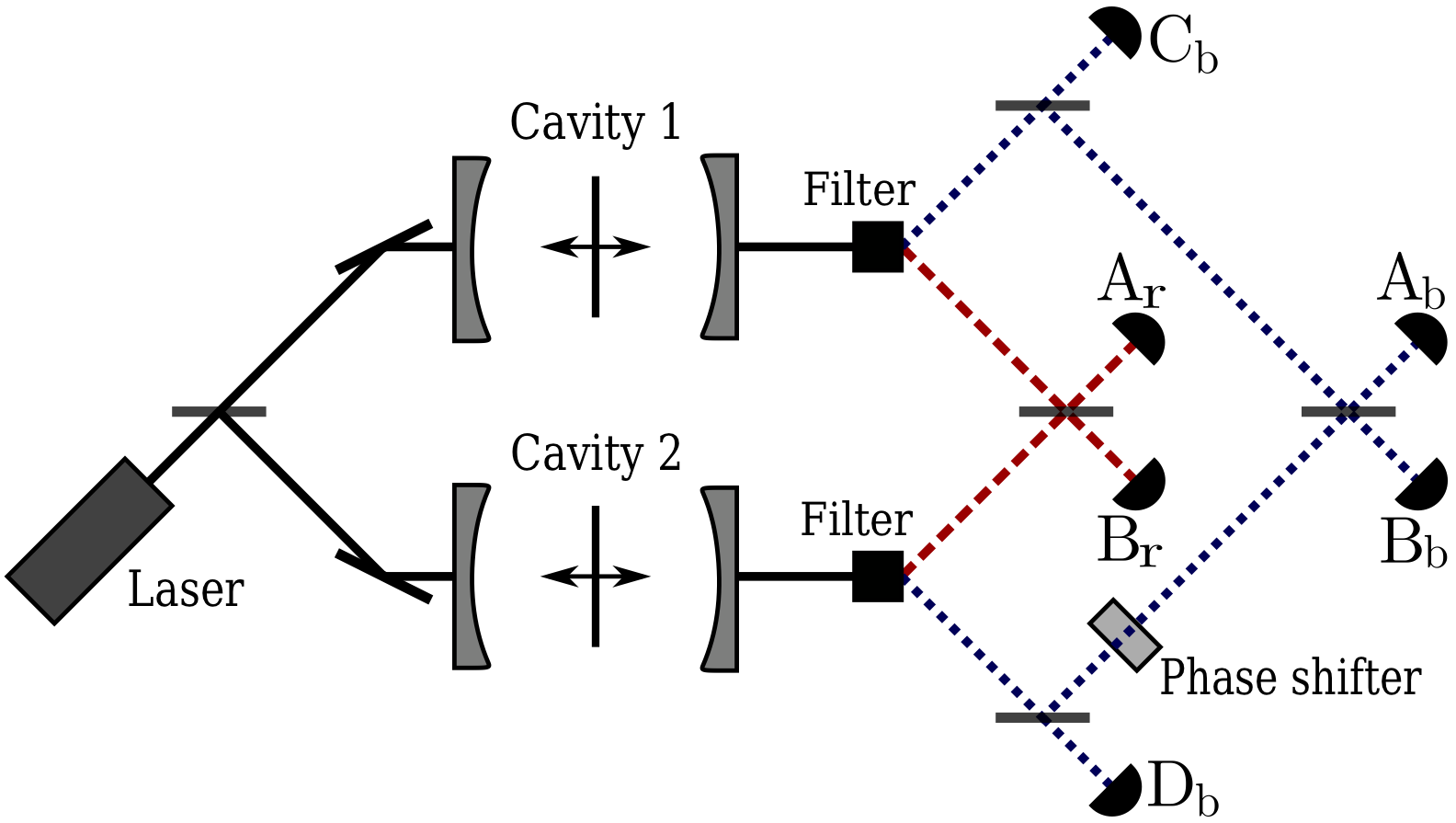}
\caption{(Color online) Schematic view of a proposed experimental setup that allows a direct measurement of $g^{(3)}_{\bb,2|\bb,1|\Ar}(\tau)$. The filter removes photons at the drive frequency and separates the sidebands into different spatial modes. Combining the red sidebands (dashed) from the two cavities on a beam splitter can create entanglement between the mechanical oscillators. The blue sidebands (dotted) are used to verify the entanglement.}
\label{fig:setupOLD}
\end{center}
\end{figure}

We start by noting that 
\begin{eqnarray}
  \label{eq:g3}
  & & \langle \bhat^\dagger_{\mathrm{b},1}(t') \bhat_{\mathrm{b},1}(t') \bhat^\dagger_{\mathrm{b},2}(t') \bhat_{\mathrm{b},2}(t') \rangle_{\Ar} \\
  & & \qquad = g^{(3)}_{\bb,2|\bb,1|\Ar}(\tau) \,  \langle \bhat^\dagger_{\mathrm{b},1}(t) \bhat_{\mathrm{b},1}(t) \rangle \langle \bhat^\dagger_{\mathrm{b},2}(t) \bhat_{\mathrm{b},2}(t) \rangle \notag \ , 
\end{eqnarray}
where we have defined the degree of third-order coherence $g^{(3)}_{k|j|i}(\tau) = \langle \bhat^\dagger_i(t) \bhat^\dagger_j(t') \bhat^\dagger_k(t') \bhat_k(t') \bhat_j(t') \bhat_i(t)  \rangle/N_{ijk}$, with $N_{ijk} = \Pi_{m = i,j,k}\langle \bhat^\dagger_m(t) \bhat_m(t) \rangle$. Also, observe that
\begin{eqnarray}
  \label{eq:coherence}
  & & \big|\langle \bhat^\dagger_{\mathrm{b},1}(t') \bhat_{\mathrm{b},2}(t') \rangle_{\Ar} \big|^2 \\
  & & \qquad  \geq \frac{1}{4} \left(g^{(2)}_{\Ab|\Ar}(\tau) - g^{(2)}_{\Bb|\Ar}(\tau) \right)^2 \langle  \bhat^\dagger_{\Ab}(t) \bhat_{\Ab}(t) \rangle^2 \notag \ .
\end{eqnarray}
Using the relation $2\langle  \bhat^\dagger_{\Ab}(t) \bhat_{\Ab}(t) \rangle = \sum_{i=1,2} \langle \bhat^\dagger_{\mathrm{b},i}(t) \bhat_{\mathrm{b},i}(t) \rangle$, 
% \begin{equation}
%   \label{eq:fluxA}
%   \langle  \bhat^\dagger_{\Ab}(t) \bhat_{\Ab}(t) \rangle = \frac{1}{2} \left(\langle \bhat^\dagger_{\mathrm{b},1}(t) \bhat_{\mathrm{b},1}(t) \rangle + \langle \bhat^\dagger_{\mathrm{b},2}(t) \bhat_{\mathrm{b},2}(t) \rangle \right) \ ,
% \end{equation}
we can derive an upper limit for $R(\tau)$ (defined in Eq.~6 of the Letter), given by
\begin{equation}
  \label{eq:Rlimit}
  R_\mathrm{m}(\tau) =  \frac{4 g^{(3)}_{\bb,2|\bb,1|\Ar}(\tau)}{\left(g^{(2)}_{\Ab|\Ar}(\tau) - g^{(2)}_{\Bb|\Ar}(\tau) \right)^2} \ .
\end{equation}
The numerator is not directly measurable in the setup in Fig.~1 of the Letter. However, one can imagine a modified setup where it is measurable, as shown here in Fig.~\ref{fig:setupOLD}. In that case, the blue and red sidebands are separated before combining the output from the two cavities. This allows a measurement of $g^{(3)}_{\Db|\Cb|\Ar}(\tau)  = g^{(3)}_{\bb,2|\bb,1|\Ar}(\tau)$, and thus an evaluation of $R_\mathrm{m}(\tau)$.
Note that for practical purposes, and to avoid detection loopholes \cite{Pearle1970PRD}, it might be better to do the measurements in two steps. The second-order coherences can be measured without the detectors $\Cb$ and $\Db$, when replacing the upper- and lowermost beam splitters with mirrors. The third-order coherence can be measured without $\Ab$ and $\Bb$, when removing the upper- and lowermost beam splitters.

The measurement of the degree of third-order coherence $g^{(3)}_{\bb,2|\bb,1|\Ar}(\tau)$ is however not necessary. The reason is that in the case of Gaussian states $\rho_\mathrm{ss}$, all higher-order correlation functions $g^{(n)}(\tau)$ with $n \geq 3$ can be expressed by degrees of second-order coherence. Specifically, we have
\begin{eqnarray}
  \label{eq:g3byg2}
  g^{(3)}_{\bb,2|\bb,1|\Ar}(\tau) & = & g^{(2)}_{\bb,1|\Ar}(\tau) + g^{(2)}_{\bb,2|\Ar}(\tau) + g^{(2)}_{\bb,2|\bb,1}(0) - 2 \notag \\
  & = & g^{(2)}_{\bb,1|\Ar}(\tau) + g^{(2)}_{\bb,2|\Ar}(\tau) - 1 \ .
\end{eqnarray}
The last equality is a result of the lack of correlations between $\bhat_{\mathrm{b},1}(t)$ and $\bhat_{\mathrm{b},2}(t)$ in the state $\rho_\mathrm{ss}$, i.e. in the absence of measurements. Assuming that the blue photon flux is the same from both cavities, i.e. $\langle \bhat^\dagger_{\mathrm{b},1}(t) \bhat_{\mathrm{b},1}(t) \rangle = \langle \bhat^\dagger_{\mathrm{b},2}(t) \bhat_{\mathrm{b},2}(t) \rangle$, we have 
\begin{equation}
  \label{eq:g212g2AB}
  g^{(2)}_{\bb,1|\Ar}(\tau) + g^{(2)}_{\bb,2|\Ar}(\tau) = g^{(2)}_{\Ab|\Ar}(\tau) + g^{(2)}_{\Bb|\Ar}(\tau)  \ .
\end{equation}
Then, Eq.~(7) of the Letter follows from Eqs.~(\ref{eq:Rlimit}), (\ref{eq:g3byg2}), and (\ref{eq:g212g2AB}).

\subsection{Relation to concurrence}
\label{sec:blablabla}
For mean phonon numbers $n_\M \ll 1$, the mechanical oscillators will for the most part occupy the two lowest energy states, $|0\rangle$ and $|1\rangle$. If we truncate to this subspace of 0 or 1 phonon, the state of the two oscillators is described by a $4 \times 4$ density matrix. Assuming that coherences between different total number of phonons can be neglected, the density matrix is
\begin{equation}
  \label{eq:rho}
 \rho = \left( \begin{array}{cccc}
p_{00} & 0 & 0 & 0 \\
0 & p_{01} & q & 0 \\
0 & q^\ast & p_{10} & 0 \\
0 & 0 & 0 & p_{11}
 \end{array} \right) \ .
\end{equation}
Here, $p_{ij}$ is the probability of having $i$ phonons in oscillator 1 and $j$ phonons in oscillator 2, whereas $q$ 
%$c(\tau) = \langle \dhat^\dagger_{2,b}(t + \tau) \dhat_{1,b}(t+\tau) \rangle_\mathrm{A,r} = \mathrm{Tr}(\dhat^\dagger_{2,b} \dhat_{1,b} \tilde{\rho}(t+\tau))$ 
is the coherence between one phonon being in either oscillator 1 or oscillator 2, given by $q = \langle \hat{c}^\dagger_{2} \hat{c}_{1} \rangle$. Since the problem is now reduced to measuring entanglement between two-level systems, we can use the concurrence as an entanglement measure \cite{Wootters1998PRL}. For the density matrix (\ref{eq:rho}), the concurrence is $\mathrm{max}(C(\tau),0)$, where 
\begin{equation}
  \label{eq:C}
  C = 2\big(|q| - \sqrt{p_{00} p_{11}}\big) \ .
\end{equation}
No entanglement gives a concurrence of 0, whereas maximal entanglement is characterized by a concurrence of 1 . 

In the weak coupling limit we have discussed, where $\tilde{\gamma} \ll \kappa$, we can relate the blue output mode $\bhat_{\bb,i}$ to the mechanical oscillator annihilation operator $\hat{c}_i$ by
\begin{eqnarray}
  \label{eq:bfromc}
  \bhat_{\bb,i}(t) & \approx & - \im \kappa_\RR \alpha \, \ex^{-\im \omega_\mathrm{D} t} \chi_\C[\omega_{\M,i}] \hat{c}_i(t)  \\ & + & \text{ vacuum noise} \ . \notag
\end{eqnarray}
This means that we can relate the entries of the density matrix (\ref{eq:rho}) to expectation values of the operators $\bhat_{\bb,i}$, $i = 1,2$, since the vacuum noise will not contribute. As discussed in the Letter, we are interested in expectation values in the state conditioned on a red photon detected at $\Ar$ at time $t$. In that non-stationary state, the entries of the density matrix become dependent on $\tau = t' - t$. The elements needed to evaluate the concurrence $C(\tau)$ are 
\begin{eqnarray}
  \label{eq:elements}
   q(\tau) & = & \frac{ \langle \bhat^\dagger_{\bb,2}(t') \bhat_{\bb,1}(t') \rangle_{\Ar}}{\left(\kappa_\RR |\alpha| |\chi_\C[\omega_{\M}]| \right)^{2}}  \ , \\ 
   p_{11}(\tau) & = & \frac{\langle \bhat^\dagger_{\mathrm{b},1}(t') \bhat_{\mathrm{b},1}(t') \bhat^\dagger_{\mathrm{b},2}(t') \bhat_{\mathrm{b},2}(t') \rangle_{\Ar}}{\left(\kappa_\RR |\alpha| |\chi_\C[\omega_{\M}]| \right)^{4} } \notag \ ,
\end{eqnarray}
and
\begin{eqnarray}
  \label{eq:qwew2}
     p_{00}(\tau) & = & 1 + p_{11}(\tau)  \\
& - & \frac{\langle \bhat^\dagger_{\bb,1}(t') \bhat_{\bb,1}(t') \rangle_{\Ar}  + \langle \bhat^\dagger_{\bb,2}(t') \bhat_{\bb,2}(t') \rangle_{\Ar} }{\left(\kappa_\RR |\alpha| |\chi_\C[\omega_{\M}]|\right)^{2} }   \ . \notag
\end{eqnarray}
For a separable state, the concurrence $C(\tau) = 0$, such that
\begin{equation}
  \label{eq:Separ}
  \frac{p_{00}(\tau) p_{11}(\tau)}{|q(\tau)|^2} \geq 1
\end{equation}
must be fulfilled. By using $p_{00} \leq 1$, Eqs.~(\ref{eq:elements}) and (\ref{eq:Separ}) produces the separability criterion presented in Eq.~(6) of the Letter. We emphasize that even though the discussion here was restricted to the subspace of 0 or 1 phonon and other assumptions, the separability criterion (6) in the Letter is of general validity.

%\bibliographystyle{abbrv}
%\bibliography{../interferometer}

\end{document}